# Machine learning reconstruction of depth-dependent thermal conductivity profile from pump-probe thermoreflectance signals

Zeyu Xiang[1], Yu Pang[1], Xin Qian[1*], Ronggui Yang[1,2*]

[1]*Department of Engineering Thermophysics, School of Energy and Power Engineering, Huazhong University of Science and Technology, Wuhan, Hubei 430074, China*

[2]*State Key Laboratory of Coal Combustion, School of Energy and Power Engineering, Huazhong University of Science and Technology, Wuhan, Hubei 430074, China*

*Corresponding emails: xinqian21@hust.edu.cn; ronggui@hust.edu.cn;

## ABSTRACT

Characterizing materials with spatially varying thermal conductivities is significant to unveil the structure-property relation for a wide range of functional materials, such as chemical-vapor-deposited diamonds, ion-irradiated materials, nuclear materials under radiation, and battery electrode materials. Although the development of thermal conductivity microscopy based on time/frequency-domain thermoreflectance (TDTR/FDTR) enabled in-plane scanning of thermal conductivity profile, measuring depth-dependent thermal conductivity remains challenging. This work proposed a machine-learning-based reconstruction method for extracting depth-dependent thermal conductivity $K(z)$ directly from frequency-domain phase signals. We demonstrated that the simple supervised-learning algorithm kernel ridge regression (KRR) can reconstruct $K(z)$ without requiring pre-knowledge about the functional form of the profile. The reconstruction method can not only accurately reproduce typical $K(z)$ distributions such as the pre-assumed exponential profile of chemical-vapor-deposited (CVD) diamonds and Gaussian profile of ion-irradiated materials, but also complex profiles artificially constructed by superimposing Gaussian, exponential, polynomial, and logarithmic functions. In addition to FDTR, the method also shows excellent performances of reconstructing $K(z)$ of ion-irradiated semiconductors from TDTR signals. This work demonstrates that combining machine learning with pump-probe thermoreflectance is an effective way for depth-dependent thermal property mapping.

**Keywords:** depth-dependent thermal conductivity, machine-learning reconstruction, frequency-domain thermoreflectance (FDTR), time-domain thermoreflectance (TDTR)

Characterizing spatially varying thermal conductivities has been increasingly significant in unveiling the structure-property relationship and the applications of functional materials[1–4]. The in-plane scanning of the thermal conductivity profile has been enabled by scanning the sample surface with local measurements, such as time- or frequency-domain thermoreflectance (TDTR/FDTR)[2,5,6] and scanning thermal microscopy (SThM)[7,8]. Mapping depth-dependent thermal conductivity profile $K(z)$ of inhomogeneous materials, however, remains an unsolved challenge, such as chemical-vapor-deposited (CVD) diamonds[9–11], ion-irradiated semiconductors near the doped or damaged sites[12–14], and electrode materials[15–17], just to name a few. Although scanning measurements could be possible by exposing a cross-section along the depth direction, sample preparation could be challenging for thin and fragile materials[18]. In addition, the spatial resolution is also limited by the sizes of the pump and probe laser spots. Recently, non-intrusive characterization of $K(z)$ is achieved using thermoreflectance measurements[11,19], leveraging the frequency dependence of thermal penetration depth $d_p = \sqrt{K/\pi C f}$, with $K$, $C$ and $f$ being the thermal conductivity, volumetric heat capacity, and modulation frequency of the pump (heating) laser, respectively. Such characterization technique requires developing thermal models that can handle nonuniform thermal conductivity by discretizing the sample into many layers, and the thermal conductivity in each layer is assumed to be uniform. To reconstruct thermal conductivity distribution from experimental data, a functional form of $K(z)$ needs to be assumed with fitting parameters, which requires pre-knowledge about the functional form of $K(z)$. However, information of $K(z)$ profile is usually not available before performing measurements, which significantly limits the application of TDTR and FDTR for characterizing novel materials with depth-dependent thermal conductivity.

In principle, the information of depth dependence of $K(z)$ is embedded in frequency-dependent thermoreflectance signals, because the thermal response at the surface is an integrated result of heat dissipation at all positions within a few penetration depths. The key question, therefore, is to make the reconstruction general enough for an arbitrary form of $K(z)$. Machine learning (ML) algorithms are particularly suitable for dealing with such complex regression tasks, because of the capability to "learn" profile shapes directly from data without pre-knowledge and human interference[20]. Recently, ML has emerged as a powerful tool for materials research[21], such as developing high-fidelity interatomic potentials[22–24], high-throughput property prediction and screening[25–30], as well as optimal structural designs[31–34]. Yet applying ML techniques to thermal characterization has been rare and limited to simple accelerations of data processing[35,36].

In this work, we demonstrate that combining simple ML regression with thermoreflectance measurements can enable robust reconstruction of depth-dependent thermal conductivity $K(z)$ without requiring any pre-knowledge, even when $K(z)$ has complicated and unknown functional forms. Our idea is briefly sketched in Fig. 1. When the thermal conductivity within a certain region deviates from the bulk value, the detected phase signals would shift accordingly when the penetration depth falls inside such region with different thermal conductivity. The machine learning model trained upon simulated $K(z)$ and frequency-domain phase signals is then applied to extract the thermal conductivity profile from the frequency-domain phase signals. The reconstruction scheme in this work is general enough for reconstructing not only simple functional forms of $K(z)$ such as linear, exponential, Gaussian, logarithmic, and so on, but also complicated thermal conductivity profiles created by superimposing these simple functions. Note that our method can also be applied to reconstruct $K(z)$ from TDTR signals, since a TDTR measurement contains thermal responses at discrete frequencies $f_n = f_0 + nf_s$ with $f_0$ being the modulation frequency and $f_s$ being the repetition rate of the pulsed laser[37]. Indeed, our reconstruction method also shows robust performance in reconstructing $K(z)$ from TDTR signals measured for ion-irradiated silicon[19].

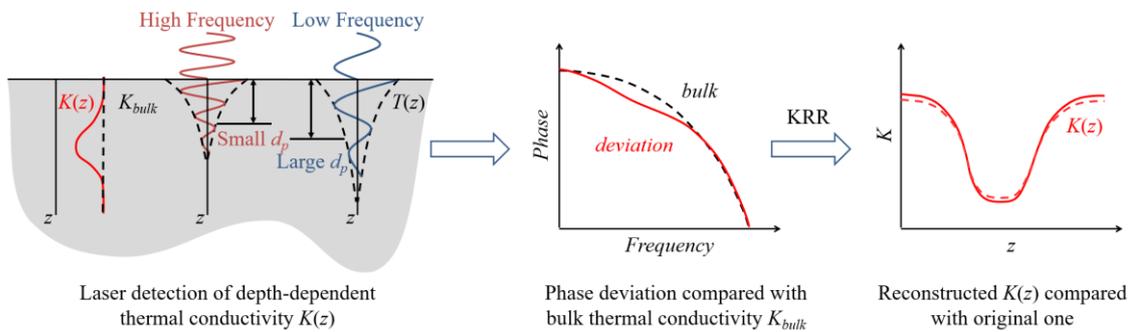

Fig. 1 Machine learning-based reconstruction of depth-dependent thermal conductivity from frequency-domain phase signals.

One of the simplest ML regression methods, kernel ridge regression (KRR), is adopted to perform the task of reconstructing $K(z)$. Although there are a variety of ML regression algorithms such as Gaussian process regression (GPR)[38], random forest (RF)[39], and neural networks (NNs)[40], this work chooses KRR[41,42] due to the following considerations. GPR and KRR are both kernel regression methods, while GPR has higher training and prediction costs[43]. RF is an ensemble learning method that lacks the transparency of training and prediction as KRR, and has higher risks of overfitting compared with KRR[39]. NNs are subject to overfitting problems with limited data[44]. As we will show in later discussions, simple KRR already showed excellent performance of reconstructing $K(z)$, even when the input frequency-domain phase signals contain a certain level of noise.

First, the ML reconstruction model needs to be trained upon different functional forms of $K(z)$ as shown in Table SI, including both typical $K(z)$ and other possible shapes. For example, CVD-grown diamonds typically have an exponential profile $K(z) = A + Be^{-Cz}$, with coefficients $A, B$ and $C$ to be fitted from experiments[11]. Close to the substrate surface with lots of nucleation sites, the grain size of CVD diamond is small and thermal conductivity increases along the cross-plane direction due to the larger grain sizes, until it reaches the bulk value away from the interface. For ion-irradiated materials, the typical profile $K(z) = \left[A + Be^{-\frac{(z-C)^2}{2D}}\right]^{-1}$ has be used considering the ion implantation concentration can be approximated as a Gaussian function[12]. Other distributions of simple functions, such as linear, parabolic, hyperbolic, and logarithmic, are also included, with controlled ranges of thermal conductivity and database shown in Table SI. In order to improve the capability of reconstructing a general thermal conductivity profile, linear combinations of functional forms are also generated,

$$K(z) = \sum_{i=1}^{n} b_i K_i(z), \tag{1}$$

where $b_i$ are random coefficients varying 0 or 1. We have generated 6×10$^4$ profiles through linear combinations of these functional forms and included them in the training and validation dataset.

For each distribution function with different combinations of coefficients, frequency-domain phase signal is computed as one input training data entry. We discretize the sample into lots of thin layers, and in each layer the thermal conductivity is regarded as uniform. Such discretization allows computation of thermal transfer matrix of a material layer with varying thermal conductivity to be simplified by a series of matrix products:

$$N(n) = M_n M_{n-1} \cdots M_1, \qquad (2)$$

where $M_j$ is the transfer matrix of the $j$-th layer located at $z_j$:

$$M_j = \begin{bmatrix} \cosh(q_j\delta) & -\frac{\sinh(q_j\delta)}{K_j q_j} \\ -K_j q_j \sinh(q_j\delta) & \cosh(q_j\delta) \end{bmatrix}, \qquad (3)$$

where $\delta = \frac{d}{n}$ is the thickness of each layer, and $q_j = \sqrt{\chi^2 + i\omega C/K_j}$ is the complex thermal wavenumber with $\chi$ being the Hankel transform variable, $K_j = K(z_j)$ being the thermal conductivity of $j$-th layer and $C$ being the volumetric heat capacity. The degree of discretization $n$ is tested until further doubling the value of $n$ would only result in 0.05% of difference in the simulated phase signal. In this work, we tested that convergence is achieved with $n = 500$ for a 5 μm thick region with varying thermal conductivity. Typical pump-probe sample geometry with a transducer layer on top of a bulk substrate is considered in this work. Hence the total transfer matrix is written as:

$$M = M_m M_G N, \qquad (4)$$

where $M_m$ is the transfer matrix of the metal transducer but evaluated with film properties such as transducer thermal conductivity $K_m$, heat capacity $C_m$, and film thickness $h_m$. $M_G = \begin{bmatrix} 1 & G^{-1} \\ 0 & 1 \end{bmatrix}$ is the matrix corresponding to the interface between the transducer and the substrate, with $G$ being the interface conductance. Control parameters including heating frequencies $f$, root-mean-square laser spot radius $w$, interface conductance value $G$, transducer and substrate properties used to compute the signals and their ranges are listed in Table SII. When generating training dataset, each combination of $K(z)$ distribution, each control parameter set $[K_m, C_m, h_m, C, G, w]$, and the corresponding phase signals are viewed as a training data point. We use Python in GPU[45] to parallelize the data generation process, and two datasets containing $3\times10^4$ and $6\times10^4$ data points are constructed for the reconstruction of depth-dependent thermal conductivity with simple functions and their linear combinations, respectively, which prove to be large enough to build a generalized reconstruction model, see later discussions.

During the training, the input vector $\boldsymbol{X}$ contains the frequency-domain phase signal in combination with the above control parameters, and the output vector $\boldsymbol{Y}$ contains $K(z)$ profiles and $G$. Choosing proper kernel functions $\mathcal{K}(\boldsymbol{X}_i, \boldsymbol{X}_j)$ is crucial for the performance of training and the reconstruction of $K(z)$. The kernel $\mathcal{K}$ is a similarity measure between the input data $X_i$ and $X_j$, expressed as $\mathcal{K}(\boldsymbol{X}_i, \boldsymbol{X}_j) = \boldsymbol{\phi}^T(\boldsymbol{X}_i)\boldsymbol{\phi}(\boldsymbol{X}_j)$, where the function $\boldsymbol{\phi}$ is referred to as feature mapping that transforms the complex nonlinear problem of reconstructing $K(z)$ into a linear one in the higher-dimensional space. Therefore, the output data is simply written as a linear combination of feature mapping or Kernel functions:

$$\boldsymbol{Y} = \boldsymbol{W}^T\boldsymbol{\phi}(\boldsymbol{X}) = \boldsymbol{\alpha}^T\boldsymbol{\mathcal{K}}, \quad (5)$$

with $\boldsymbol{W}^T$ being the matrix of linear coefficients, $\boldsymbol{\mathcal{K}}$ being the kernel matrix whose elements are $\mathcal{K}(\boldsymbol{X}_i, \boldsymbol{X}_j)$, $\boldsymbol{\alpha}$ being the linear coefficient vector of kernels. When training the KRR model, feature mapping $\boldsymbol{\phi}$ and the coefficients $\boldsymbol{W}$ do not explicitly involve numerical computations[46]. Instead, the coefficient of kernel functions $\boldsymbol{\alpha}$ is determined by inverting the regularized kernel matrix: $\boldsymbol{\alpha} = (\boldsymbol{\mathcal{K}} + \lambda \boldsymbol{I})^{-1}\boldsymbol{Y}$, where $\boldsymbol{I}$ is the identity matrix and $\lambda$ is the regularization factor. After the optimal coefficient $\boldsymbol{\alpha}$ is obtained, the reconstruction is performed by first computing kernel functions between the measured signal and training signals, and the thermal conductivity profile is then evaluated numerically using Eq. (5). Possible choices of kernel functions include Gaussian functions, polynomials, and Laplacian functions, together with the associated hyperparameters, are listed in Table SIII in Supplementary Material. In order to evaluate the regression performance of different kernel functions, the coefficient of determination ($R^2$) is defined as the index for evaluating regression accuracy:

$$R^2 = 1 - \frac{(Y-\hat{Y})^2}{(Y-\bar{Y})^2}, \quad (6)$$

where $\hat{Y}$ and $\bar{Y}$ are the predicted and the average thermal conductivity distributions, respectively. Better regression performance corresponds to larger $R^2$ values (closer to 1). The optimal kernel functions, regularization factors and corresponding hyperparameters are determined through grid-search cross-validation implemented by Skit-learn[47], where all the possible combinations are validated by $k$-fold cross-validation. Fig. 2a shows the workflow of $k$-fold cross-validation. The dataset is first randomly divided into $k$ subsets, with only $(k-1)$ subsets used for training, and the remaining subset for validation. The performance of certain kernel functions and hyperparameters is denoted as $R_j^2$ if the $j$-th subset is used for validation and the rest subsets are training subsets. Similar processes are repeated $k$-times until each dataset has been validated once. The $k$-fold cross-validated $R^2$ is calculated by averaging among all $R_j^2$:

$$R^2 = \frac{1}{k}\sum_{j=1}^{k} R_j^2. \quad (7)$$

This work uses five-fold cross-validation ($k=5$) to evaluate model performance, since further increasing $k$ would result in a negligibly small change in optimal hyperparameters of the kernel functions. Fig. 2b shows the cross-validated $R^2$ of typical kernel functions, and clearly the Laplacian $\mathcal{K}(X_i, X_j) = \exp(-\beta \parallel X_i - X_j \parallel)$ has the highest $R^2$, which is therefore chosen for developing the KRR model for $K(z)$ reconstruction.

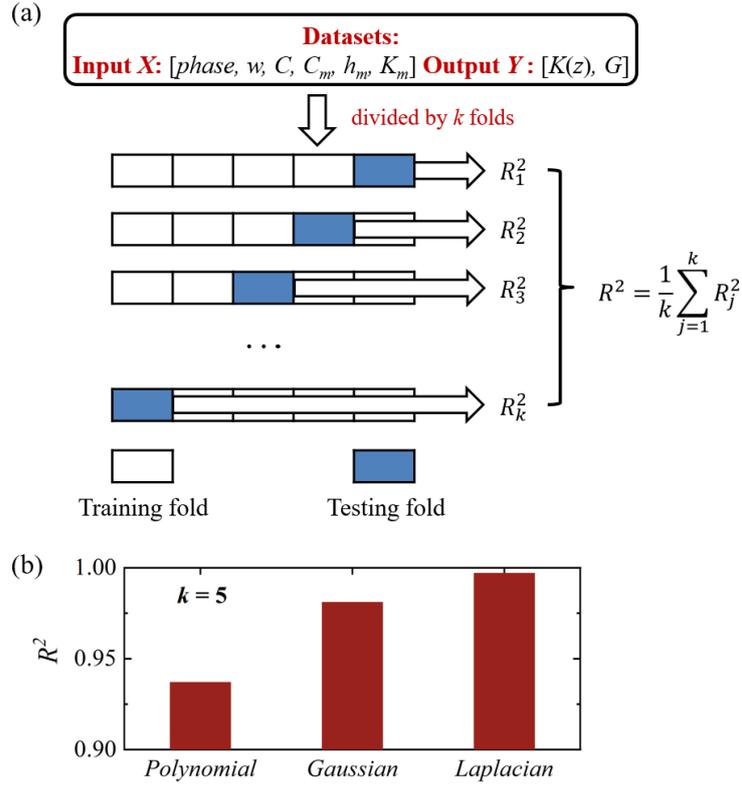

Fig. 2 (a) Workflow of $k$-fold cross-validation and (b) coefficients of determination $R^2$ for different choices of kernel functions.

Experimental noise is another factor that needs to be considered when reconstructing depth-dependent thermal conductivity. A random fluctuation with a normal distribution ranging from -0.5% to 0.5% is added to the phase signal to mimic the experimental noise with a typical signal-to-noise ratio of 200, of which the effect on reconstructed thermal conductivity profile and interface conductance is presented in Fig. S1 of Supplementary Materials. For the reconstruction of $3\times10^4$ profiles of $K(z)$ with simple functions, the trained KRR performs well, of which the determination coefficient $R^2$ is 0.997 tested by another group of $1\times10^4$ datasets. Fig. 3 shows the performance of KRR-reconstructed $K(z)$ of the most two typical samples according to their phase deviations from those of uniform ones (Fig. 3a-b), like CVD diamonds with exponential thermal conductivity profile (Fig. 3c) and the ion-irradiated material with Gaussian-like thermal resistivity (Fig. 3d). Excellent agreement has been achieved with relative deviations smaller than 1%. Due to the high dimensionality of coefficients $\alpha$ of kernel functions, error propagation using simple contour methods designed for low dimensional empirical fitting is no longer feasible. Therefore, Monte Carlo method[35] is used in this work to statistically compute the distributions of unknown parameters. Typical uncertainties ($2\sigma$, 95% confidence level) of control parameters are, 3% for $C_m$ and $C$, 4% for $w$, 5% for $h_m$, and 10% for $K_m$[48,49]. Fig. 3e-f present the resulting uncertainties of KRR-reconstructed exponential thermal conductivity and Gaussian-like thermal resistivity, which are smaller than 10% and 20% in the whole range, respectively. The performance of reconstructing other simple $K(z)$ functions is also shown Supplementary Materials S4, indicating that KRR is robust enough for high-fidelity reconstructions even with the disturbance of noises.

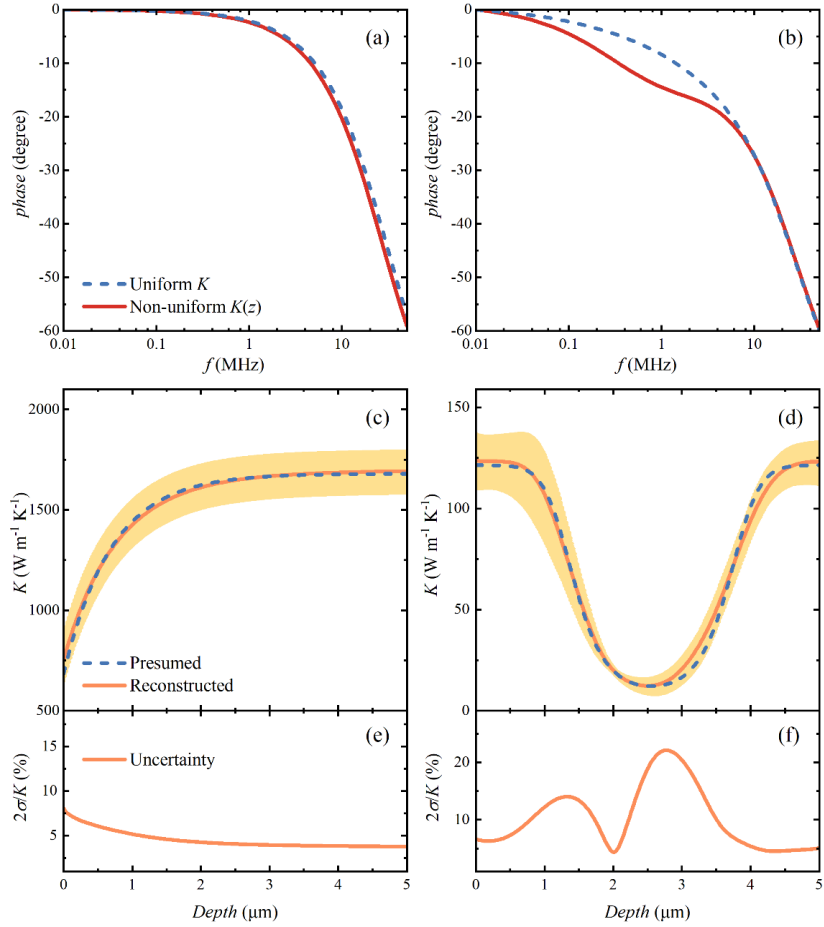

Fig. 3 (a-b) Phase deviations of depth-dependent thermal conductivity profiles (solid line) from those of uniform ones (dash line) of (a) exponential distribution and (b) Gaussian distribution. (c-d) Depth-dependent thermal conductivity profiles (dash line) and the reconstructed ones (solid line) with uncertainties (shadowed area) of (c) exponential distribution and (d) Gaussian distribution. (e-f) Uncertainties of thermal conductivity with (e) exponential distribution and (f) Gaussian distribution.

Besides simple $K(z)$ profiles, the ML model can also reconstruct complicated functional forms of $K(z)$ generated by randomly superimposing those simple distribution functions. The KRR model trained upon $6\times10^4$ complicated $K(z)$ profiles showed a high determination coefficient $R^2$ of 0.985, which is tested upon $2\times10^4$ $K(z)$ profiles. Fig. 4a-c show several random combinations of the above thermal conductivity distribution profiles and the reconstructed ones, with uncertainties lower than 15%. By superimposing simple functional shapes, curvature changes and even oscillations in the $K(z)$ profile can be generated, while the KRR-reconstruction model developed in this work can still capture these features, clearly showing that machine learning method is capable of reconstructing arbitrary distributions of $K(z)$ without any pre-knowledge. Fig. 4d further demonstrates robust performance of reconstructing periodically modulated $K(z)$ which is typical for superlattices. The ML model can also extract thermal conductivity profile near an intense defect layer, which is shown in Fig. S2 and Fig. S3 of Supplementary Materials.

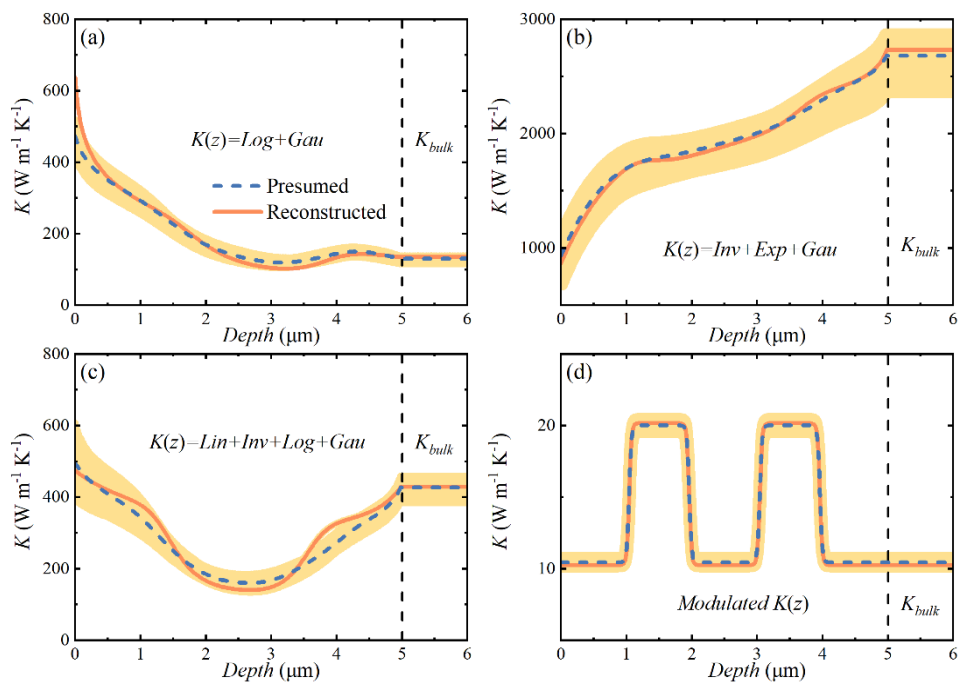

Fig. 4 Presumed and reconstructed depth-dependent thermal conductivity of complicated profiles generated by linear combinations of (a) logarithmic and Gaussian (*Log+Gau*) distributions, (b) hyperbolic, exponential, and Gaussian (*Hyp+Exp+Gau*) distributions, (c) linear, hyperbolic, logarithmic, and Gaussian (*Lin+Hyp+Log+Gau*) distributions, (d) square-wave-like distributions. Shadowed area represents uncertainty bounds.

In addition to reconstructing $K(z)$ using simulated FDTR phase signals, we also validated the capability of reconstructing $K(z)$ from experimental TDTR signals measured at 2.1 MHz upon ion-irradiated silicon by Pfeifer et al.[19] As shown in Fig. 5, the trained KRR model showed excellent agreement with the empirical fitting when the depth $z < 1$ μm. The thermal conductivity in the region without irradiations is ~ 10% higher than reported by Pfeifer et al.[19], but such deviation is within typical uncertainty level of TDTR measurement[48]. One possible reason for such deviation could be attributed to the fact that frequency components embedded in TDTR signal are discrete. The lowest frequency component corresponds to the modulation frequency 2.1 MHz, and the penetration depth is estimated around 3 μm. The second lowest frequency, however, is as high as 82.1 MHz with a repetition rate of 80 MHz, and the penetration depth is only 0.5 μm. As a result, TDTR lacks the capability to continuously sample responses at different depths, which would give rise to relatively larger uncertainty of reconstructing $K(z)$ at larger depths.

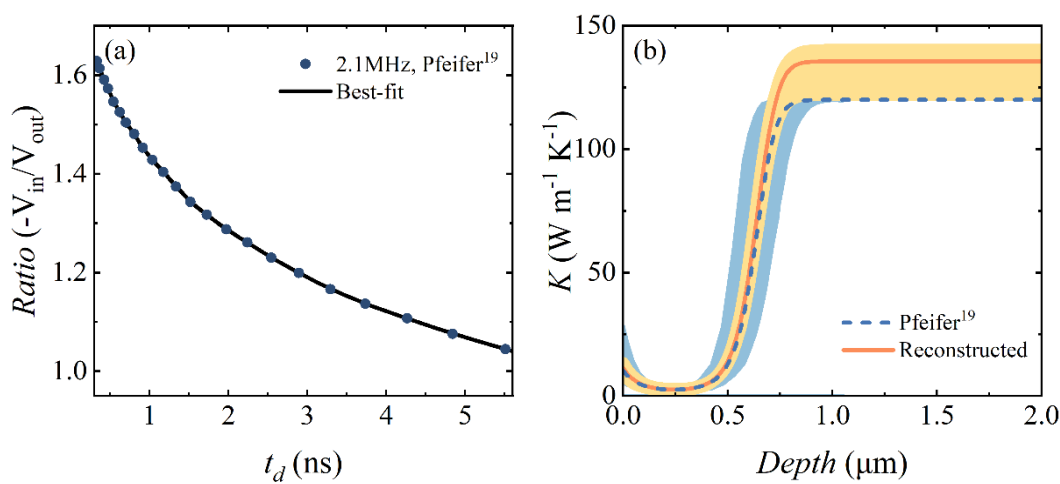

Fig. 5 Validation using experimental TDTR signals.

Finally, we comment on the working range of thermal conductivity profile reconstruction. First, the region of varied thermal conductivity must be accessible by the thermal penetration so that the phase deviation is measurable. Typical modulation frequency of FDTR varies from 0.01 MHz to 50 MHz, corresponding to a tunable range of nearly two-orders of magnitude in thermal penetration depth $d_p$, namely the maximum reconstruction depth. Depending on the materials with different thermal diffusivity, $d_p$ can vary from a few to nearly hundreds of micrometers. In Supplementary Material S4-5, reconstruction of hypothetical inhomogeneous materials with a wide range of thermal conductivity is also performed. Typical relative uncertainty of the reconstructed $K(z)$ is less than 20%, thus relative change of $K(z)$ smaller than 20% of the mean value can no longer be reconstructed using the proposed method.

To conclude, this work leveraged ML methods for thermoreflectance measurements and achieved high-fidelity reconstruction of depth-dependent thermal conductivity in complex functional materials, without requiring pre-knowledge about the functional forms of thermal conductivity distribution. Using simple KRR regression method combined with grid-search cross-validation techniques, the ML-based reconstruction method can extract both profiles described by simple functions, as well as complex thermal conductivity distribution with curvature changes and oscillations. After proper training, the reconstruction method can output thermal conductivity profile $K(z)$ directly from the frequency-domain phase signals without expensive computations of the discretized thermal model. This work shows that machine learning could be an effective way for mapping spatially varying thermal properties. Although we focused on isotropic thermal conductivity, it is possible to simultaneously determine depth-dependent cross-plane thermal conductivity and anisotropy ratio by combining machine learning algorithms with variable spot-size FDTR or TDTR measurements, which would require expanding training dataset and implementation of more efficient reconstruction algorithms in the future.

**Acknowledgment**

R.Y. acknowledges funding support from National Key Research and Development Program of China (Grant No. 2022YFB3803900). X. Q. acknowledges support from National Natural Science Foundation of China (NSFC Grant No. 52276065) and start-up funding support by Huazhong University of Science and Technology.

# Supplementary Materials: Machine learning reconstruction of depth-dependent thermal conductivity profile from pump-probe thermoreflectance signals

Zeyu Xiang[1], Yu Pang[1], Xin Qian[1*], Ronggui Yang[1,2*]

[1]*Department of Engineering Thermophysics, School of Energy and Power Engineering,*

*Huazhong University of Science and Technology, Wuhan, Hubei 430074, China*

[2]*State Key Laboratory of Coal Combustion, School of Energy and Power Engineering,*

*Huazhong University of Science and Technology, Wuhan, Hubei 430074, China*

*\*Corresponding authors: xinqian21@hust.edu.cn; ronggui@hust.edu.cn;*

# Contents

# S1. Training datasets and kernel functions

**TABLE SI.** Functional forms of depth-dependent thermal conductivity

| Functionals | Formula $K(z)$ | Ranges (W m$^{-1}$ K$^{-1}$) | Profile numbers |
|---|---|---|---|
| Linear | $A + Bz$ | 1-2000 | 5000 |
| Hyperbolic | $(A + Bz)^{-1}$ | 1-2000 | 5000 |
| Logarithmic | $\ln(A + Bz)$ | 1-2000 | 5000 |
| Polynomial | $A + Bz + Cz^2$ | 1-2000 | 5000 |
| Exponential | $A + Be^{-Cz}$ | 1-2000 | 5000 |
| Gaussian | $\left[A + Be^{-\frac{(z-C)^2}{2D}}\right]^{-1}$ | 1-2000 | 5000 |
| Modulated | Eq. (S1) | 1-200 | 5000 |

To imitate the modulation of in some structures which have a modulated thermal conductivity profile, we first try the-square-wave-like distributions, which can be decomposed to a series of sine functions. For the continuity of $K(z)$ over the whole depth range (5 µm), a variation of sigmoid function (Eq. (S1)) is adopted to generate the square-wave-like profile. The period T ranges from 1 µm to 2 µm, which is realized by artificial inversion symmetry.

$$K(z) = \begin{matrix} A + [B + e^{-Cz}]^{-1}, & z < T/2 \\ A + [B + e^{Cz}]^{-1}, & z > T/2 \end{matrix}, \qquad (S1)$$

**TABLE SII.** The range of control parameters used in the training database.

| Parameters | Ranges | Unit |
|---|---|---|
| $K_m$ | 100-250 | W m$^{-1}$ K$^{-1}$ |
| $C_m$ | 2.3-2.6 | MJ m$^{-3}$ K$^{-1}$ |
| $h_m$ | 80-120 | nm |
| $C$ | 1.0-3.0 | MJ m$^{-3}$ K$^{-1}$ |
| $G$ | 30-50 | MW m$^{-2}$ K$^{-1}$ |
| $w$ | 2.5-3.5 | μm |
| $f$ | 0.01-50 | MHz |

**TABLE SIII.** Typical kernel functions and the expressions.

| Kernel functions | Expressions | Hyperparameters |
|---|---|---|
| Gaussian | $\mathcal{K}(X_i, X_j) = \exp(-\beta \parallel X_i - X_j \parallel^2)$ | $\beta$ |
| Laplacian | $\mathcal{K}(X_i, X_j) = \exp(-\beta \parallel X_i - X_j \parallel)$ | $\beta$ |
| Polynomial | $\mathcal{K}(X_i, X_j) = (\beta X_i X_j + c)^d$ | $\beta$, $c$, $d$ |

For finding optimal kernel functions of reconstructing $K(z)$ with both simple functions and their combinations, kernel functions among Gaussian, Laplacian, and polynomial and their hyperparameters $\beta$ in the range of 0.01-10, $d$ in the range of 2-5, and $c$ valued as 1 are verified through grid-search cross-validation. The found optimal kernel functions are Laplacian, with $\beta$ being 0.135 and 0.129 for reconstructing $K(z)$ with simple functions and their combinations, respectively.

## S2. $K(z)$ reconstruction using signals with noise

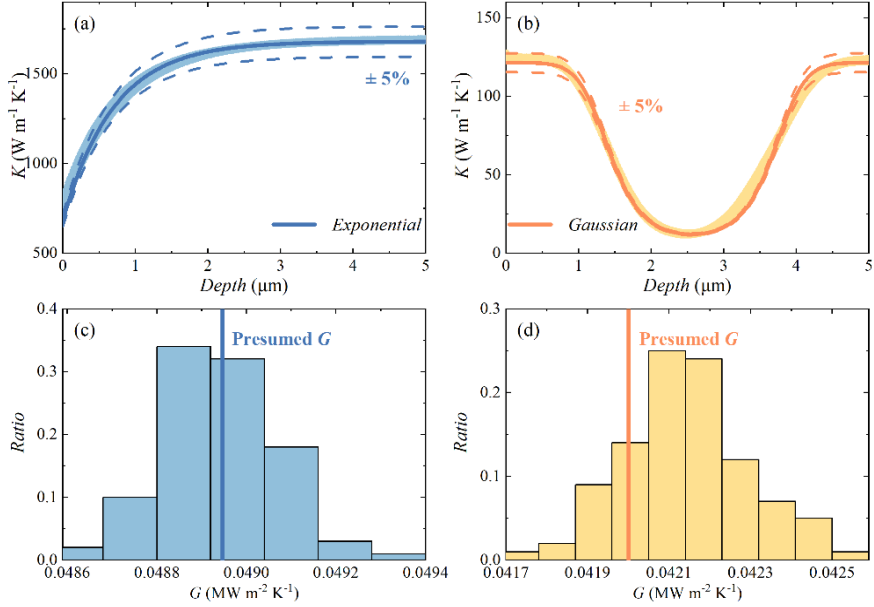

**Fig. S1** (a-b) Depth-dependent thermal conductivity profiles (solid line) and the reconstructed ones (shadowed area) under noise (signal-to-noise ratio of 200) of (a) exponential distribution and (b) gaussian distribution. (c-d) Interfacial thermal resistance (solid line) and the reconstructed ones (histograms) of (c) exponential distribution and (d) gaussian distribution.

## S3. $K(z)$ reconstruction near an intense defect layer

We test the performance of reconstructing $K(z)$ near an intense defect layer by reconstructing a Gaussian dip in thermal conductivity profile (Fig. S2). We found that the performance is affected by the full-width-half maximum (FWHM) of the profile. When the FWHM is larger than 0.2 μm, the reconstruction performance is reasonable, but our method fails when FWHM is only 0.1 μm, which indicates the limit of cross-plane spatial resolution of our method.

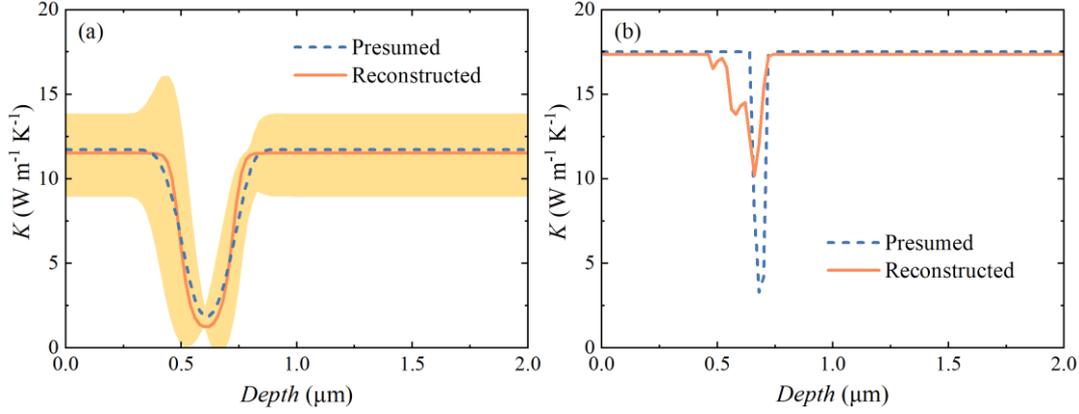

**Fig. S2** Depth-dependent thermal conductivity profiles (dash line) and the reconstructed ones (solid line) with uncertainties (shadowed area) of discrete layers or intense defects with full-width half maxima of (a) 0.3 μm, (b) 0.1 μm.

A possible extreme is the defect layer is thin enough such that the varying thermal conductivity can be lumped into a single interface resistance. In this case, we can directly use the transfer matrix method to solve the heat equations without machine learning algorithms, by setting the interface conductance and its position as fitting parameters. We find it is possible to simultaneously determine these two parameters with relative uncertainties less than 10%, as shown by the Monte Carlo histograms in Fig. S3.

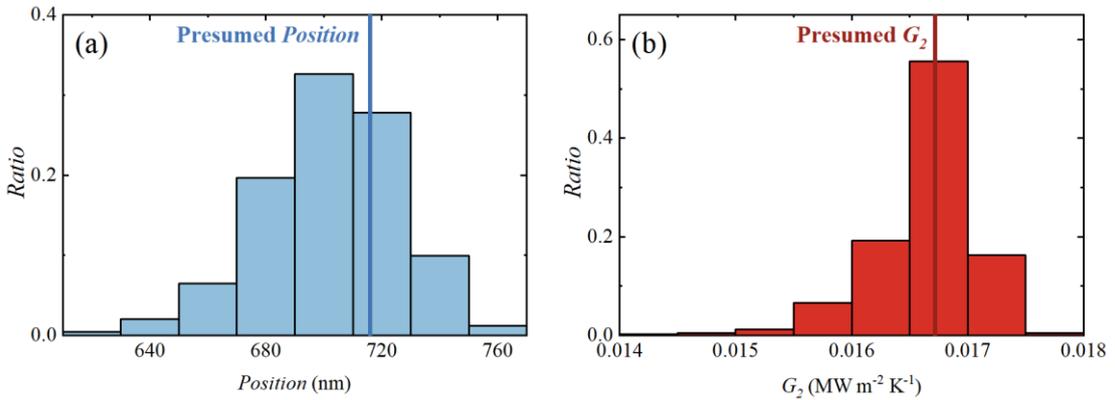

**Fig. S3** (a) Presumed positions of $G_2$ (solid line) and the reconstructed ones with uncertainties (histograms). (b) Presumed values of $G_2$ (solid line) and the reconstructed ones with uncertainties (histograms).

## S4. Uncertainty of materials with high $K(z)$ ($K_{max} > 100$ W m$^{-1}$ K$^{-1}$)

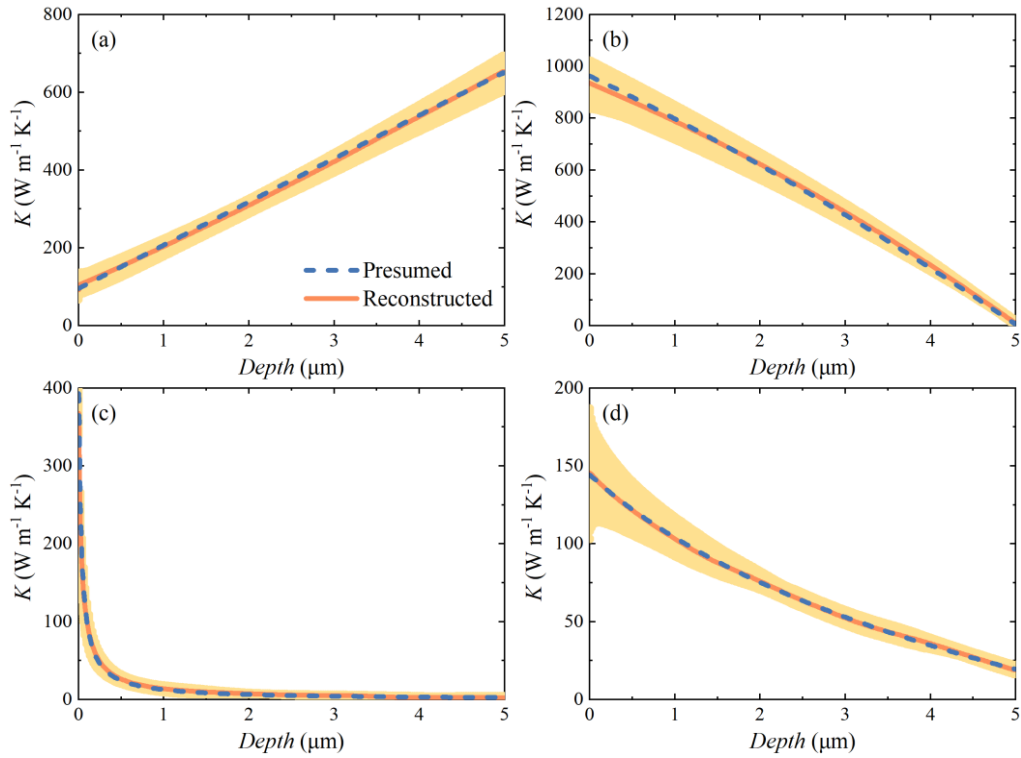

**Fig. S4** (a-d) Depth-dependent thermal conductivity profiles (dash line) and the reconstructed ones (solid line) with uncertainties (shadowed area) of (a) linear distribution, (b) parabolic distribution, (c) hyperbolic distribution, and (d) logarithmic distribution.

## S5. Uncertainty of materials with low $K(z)$ ($K_{max} < 2$ W m$^{-1}$ K$^{-1}$)

Excellent reconstructions of different $K(z)$ profiles are realized with relative deviations smaller than 1% from the whole picture. Besides, the reconstruction uncertainties of different $K(z)$ profiles are smaller than 10% in the whole range, except for the 1.5-2.0 μm range of logarithmic and exponential distributions (<15%) and the 0.5-1 μm range of Gaussian distribution (<30%).

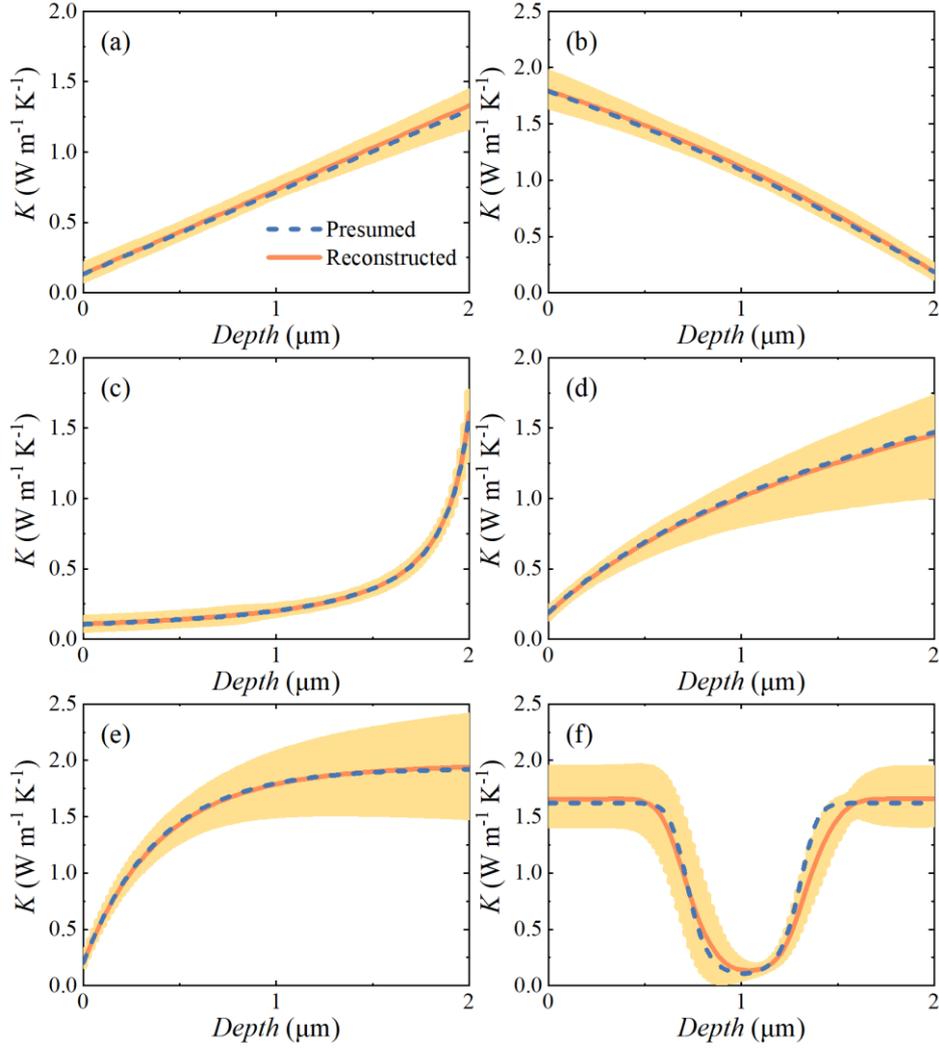

**Fig. S5** (a-f) Depth-dependent thermal conductivity profiles (dash line) and the reconstructed ones (solid line) with uncertainties (shadowed area) of (a) linear distribution, (b) parabolic distribution, (c) hyperbolic distribution, (d) logarithmic distribution, (e) exponential distribution, and (f) Gaussian distribution.

Excellent reconstructions of complex $K(z)$ profiles are realized with relative deviations smaller than 2% from the whole picture. Besides, the reconstruction uncertainties of different $K(z)$ profiles are smaller than 20% in the whole range.

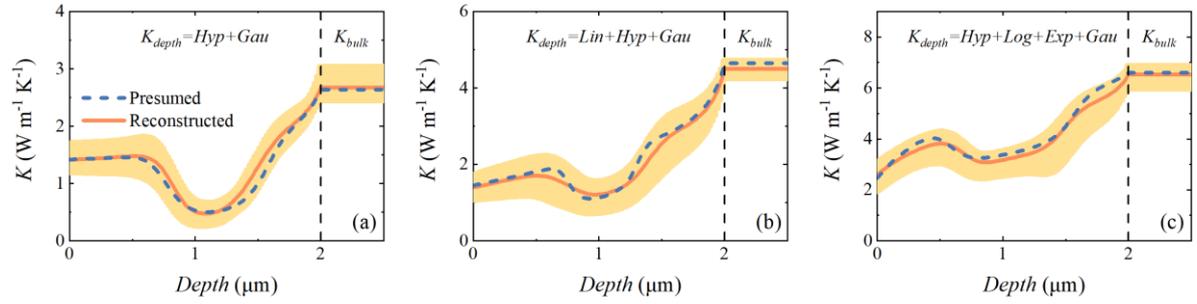

**Fig. S6** (a-c) Depth-dependent thermal conductivity profiles (dash line) and the reconstructed ones (solid line) with uncertainties (shadowed area) of a combination of (a) hyperbolic and Gaussian (Hyp+Gau) distributions, (b) linear, hyperbolic, and exponential (Lin+Hyp+Gau) distributions, (c) hyperbolic, logarithmic, exponential, and Gaussian (Hyp+Log+Exp+Gau) distributions.